\documentclass[3p,times,twocolumn]{elsarticle}

\usepackage{ecrc}

\volume{00}

\firstpage{1}

\journalname{Nuclear Physics B Proceedings Supplement}

\runauth{A. Marcowith et al.}

\jid{nuphbp}

\jnltitlelogo{Nuclear Physics B Proceedings Supplement}

\usepackage{amssymb}

\usepackage[figuresright]{rotating}

\begin{document}

\begin{frontmatter}




\title{Cosmic-ray acceleration and gamma-ray signals from radio supernov\ae}

 \author[label1]{A. Marcowith}
 \author[label1]{M. Renaud}
 \author[label2]{V. Dwarkadas}
 \author[label3]{V. Tatischeff}
 \address[label1]{Laboratoire Univers et particules de Montpellier, Universit\'e Montpellier II/CNRS, place E. Bataillon, cc072, 34095 Montpellier, France}
 \address[label2]{Department of Astronomy and Astrophysics, University of Chicago, Chicago, Illinois, 60637, USA}
 \address[label3]{Centre de Sciences Nucl\'eaires et de Sciences de la Mati\`ere, IN2P3/CNRS and Univ Paris-Sud, 91405 Orsay, France}
 

\begin{abstract}
Core collapse supernovae (SNe) are among the most extreme events in
the universe. The are known to harbor among the fastest (but non- or
mildly-relativistic) shock waves. Once it has crossed the stellar atmosphere,
the SN blast wave expands in the wind of the massive star progenitor. 
In type IIb SNe, the progenitor is likely a Red SuperGiant (RSG)
star which has a large mass loss rate and a slow stellar wind
producing a very dense circumstellar medium. A high velocity shock and
a high density medium are both key ingredients to initiate fast
particle acceleration, and fast growing instabilities driven by the
acceleration process itself. We have reanalyzed the efficiency of
particle acceleration at the forward shock right after the SN
outburst for the particular case of the well-known SN 1993J. We find that 
plasma instabilities driven by the energetic particles accelerated at the
shock front grow over intraday timescales. This growth, and the interplay 
of non-linear process, permit a fast amplification of the magnetic field at 
the shock, that can explain the magnetic field strengths deduced from the 
radio monitoring of the source. The maximum particle energy is found to reach 
1-10 PeV depending on the instability dominating the amplification process. 
We derive the time dependent particle spectra and the associated hadronic
signatures of secondary particles (gamma-ray, leptons and neutrinos) arising
from proton proton interactions. We find that the Cherenkov Telescope Array (CTA) should easily detect objects 
like SN 1993J in particular above 1 TeV, while current generation of Cherenkov 
telescopes such as H.E.S.S. could only marginally detect such events. The 
gamma-ray signal is found to be heavily absorbed by pair production process 
during the first week after the outburst. We predict a low neutrino flux above 
10 TeV, implying a detectability horizon with a KM3NeT-type telescope of 1 Mpc
only. We finally discuss the essential parameters that control the particle 
acceleration and gamma-ray emission in other type of SNe.
\end{abstract}

\end{frontmatter}


\section{Introduction}
\label{S:Int}

It is widely accepted that SuperNova Remnants (SNRs) are likely the sources 
of Cosmic Rays (CR) in our Galaxy up to energies corresponding
to the so-called CR ``knee'' at 3 PeV for the proton component (e.g. see
\cite{Blasi13}). However, an important question remains: if SNRs do
produce PeV particles, when does this acceleration occur in the SNR
lifetime ? Another issue is that in order to explain the whole CR
spectrum one has to consider particles to be accelerated not only up
the knee but up to a few hundred PeV, where an extragalactic component
should start taking over. Hence, only a restricted number of Galactic sources
should be able to produce such energetic particles following the
confinement criterium (see \cite{Hillas84}) stating that the Larmor
radius of the particle should match the source size. Either
extended sources with standard interstellar medium magnetic field
values or more compact sources but with more intense magnetic fields
could fulfill such a criterium. In the first category one finds the
massive star clusters and their superbubbles \cite{Bykov01,
 Parizot04, Ferrand10} while the second category comprises young
supernova remnants \cite{Bell13}. This latter work has already pointed
out the possibility to produce multi-PeV CR energies in very young SNRs
rather than in more evolved objects entering the Sedov phase. It is
then very important to propose accurate modeling of the time history
of CR acceleration in a SNR that can be tested by forthcoming instruments 
such as the Cherenkov Telescope Array (CTA). \\

This work addresses the issue of particle acceleration and gamma-ray
radiation in the so-called wind SNe arising from massive star progenitors
and evolving in the dense CSM. It provides observationally testable modeling 
of these early stages. The novelty here is that we consider particle 
acceleration right after the outburst when the radio luminosity is close
to its maximum. The discussion in this paper is mainly focused on one 
particular well-monitored object: SN 1993J. We will review the main results 
of radio observations and their modeling in \S \ref{S:Acc}. Some perspectives 
for other types of SNe will be discussed in \S \ref{S:Dis}. We have adapted 
recent models of magnetic field amplification to the case of SN 1993J in \S
\ref{Su:MFA}. High magnetic field means likely higher CR energies as
stated above. This will be examined in \S \ref{S:Max}. We will detail
the modeling of the different particle distribution species in radio
SNe in \S \ref{S:CRs}. Once the particle distribution has been derived
we will address the multi-wavelength radiation (in \S \ref{S:MWs}) and
the neutrino (in \S \ref{S:NEu}) signal that would have been detected
from SN 1993J by modern instruments if they were operational at the
time of the SN outburst. Our main prediction concerning the gamma-ray
radiation is detailed in \S \ref{S:GRa} and a conclusion is given in 
\S \ref{S:Con}.

\section{A case study: SN 1993J}
\label{S:Acc}
Our model is based on the work proposed by \cite{Tatischeff09} (hereafter
T09) to account for particle acceleration in SN 1993J. This SN is a type IIb 
SN resulting from the explosion of a massive star with an initial mass between 
13 to 20 $M_{\odot}$ (in the case the explosion
occurred in a binary system) or 20 to 30 $M_{\odot}$ in the case of an
isolated star \cite{Maund04}. The star then evolved into a Red
SuperGiant (RSG) phase with a mass loss rate of $\sim 3.8 \times 10^{-5}
M_{\odot}/{\rm yr}$ and a slow wind velocity of $u_{w} \sim 10 \ \rm{km/s}$ 
(see T09 for the derivation of these parameters). These values imply that in
case of a constant mass loss rate the density in circumstellar medium
(CSM) scales as:
\begin{equation}
n_{circ}= {\dot{M} (1+2X) \over 4\pi r^2 u_w m_H (1+4X)} \ ,
\end{equation}
where $X=0.1$ is the Helium fraction and $m_H$ is the hydrogen atom
mass. With the above parameters the effective density downstream of the
forward shock is typically $n_{eff} \simeq 4\times 10^9
\ \rm{cm^{-3}}$ at the time of outburst (here we have accounted for a
shock compression ratio of 4). The radius of the shock at the outburst
$r \simeq 3.5\times 10^{14} \ \rm{cm}$ (see T09) has been deduced from
the derivation of the radio expansion. The shock is expected to
propagate into a fully ionized medium \citep{Fransson96}.

T09 proposed a scenario based on the radio follow-up observations of
SN 1993J over a period covering 3100 days after the SN outburst. The
synchrotron model provides fits to light curves at 6 different radio
wavelengths, including the effects of synchrotron-self-absorption and
free-free absorption. The effect on the radio signal attenuation by
the homogeneous CSM matter, as well as the presence of clumps, has
been taken into account in the modeling. This resulted in the
following time dependence of the averaged total magnetic field from
the synchrotron emitting shell:
\begin{equation}
\label{Eq:MFo}
\langle B \rangle \simeq [2.4 \pm 1 \ \rm{G} ] \times \left( {t \over 100~\rm{days}} \right)^{-1.16 \pm 0.20} \ ,
\end{equation}
which points towards a magnetic field of the order of $500$ G after 1
day. This field is consistent with observations \cite{Chandra04}. 
The above magnetic field is only an averaged value
over the synchrotron shell and does not represent the magnetic field
produced at the forward shock necessarily. In the context of the model
of CR driven instabilities detailed afterward, we will identify this
field with the one obtained in the post shock gas of the forward shock
front. The magnetic field in the upstream medium can hence be derived
from the generalized Rankine-Hugoniot conditions at the shock front
accounting for the CR back-reaction and the turbulent magnetic field
heating at the shock precursor (see details in \cite{Parizot+06, Marcowith10}).

\subsection{Magnetic field amplification (MFA)}
\label{Su:MFA}
Eq.~(\ref{Eq:MFo}) clearly shows that the magnetic field is still much
larger than typical field expected in the wind of massive stars at
such distance. For comparison the equipartition magnetic field in the
progenitor wind is \citep{Fransson98}:
\begin{eqnarray}
\label{Eq:Beq}
B_{eq} &=& {\left(\dot{M} u_{w}\right)^{1/2} \over r} \nonumber \\
& &\simeq [2.5 \ \rm{mG}] \times \dot{M}_{-5}^{1/2} \times u_{w,10}^{1/2} \times r_{16}^{-1} \ ,
\end{eqnarray}
about one thousand times less than the value derived in
Eq.\ref{Eq:MFo}. The solutions of particle acceleration by non-linear
diffusive shock acceleration model (based on the approach of
\cite{Berezhko99}) proposed in T09 are not strongly modified by the CR
pressure. This means that the sub-shock compression ratio should not
be far from 4. Then the compression of the magnetic field given by
Eq.~(\ref{Eq:Beq}) cannot explain the field value given by
Eq.~(\ref{Eq:MFo}) (see also the discussion in \cite{Fransson98}). {\it
  We conclude and assume from now that a strong magnetic field
  amplification process is at work at the forward shock front and it
  is driven by the diffusive shock acceleration (DSA) of hadrons.} \\

We now discuss in this framework the possible mechanisms that can
produce such an amplification. Note that for our purpose it is simply
required that there exists a fast instability to produce the requested
MFA at scales where the CRs are efficiently scattered. The magnetic
field amplification at fast shocks in young SNR is likely connected
with the presence of energetic particles (likely to become CRs) moving
in front of the shock and producing a precursor. We can make a
distinction among several types of instabilities: the instabilities
produced by the pressure gradient of cosmic rays, the instabilities
produced by the streaming of cosmic rays (see \cite{Bykov12} for a
recent review). The magnetic field can be amplified due to the
presence of background turbulence, and further amplified at the shock
front.

\subsubsection{Streaming driven instabilities} 

The streaming of cosmic rays ahead the shock front produces magnetic
fluctuations that have two different different regimes. The streaming
modes can either be in resonance (R) with the energetic particles,
i.e. in the high-energy limit they have a wave-number such that $k
\sim \rm{r}_L^{-1}$ or they can be non-resonant (NR) with a
wave-number much larger than $\rm{r}_L^{-1}$ \cite{Bell00, Bell04,
  Amato09}. The NR modes grow the fastest especially in the regime of
fast shocks (see \cite{Pelletier+06}). As already discussed in T09
within the conditions that prevails for IIb SN, the NR modes grow and
produce magnetic fluctuations over intra-day timescales; we get from
the time of the outburst using the shock parameters described above a
NR mode growth time of:
\begin{eqnarray}
\label{Eq:TNR}
\tau_{NR-st} &= &[0.16~\rm{day}] \times \left({\phi/15 \over (\xi_{CR}/0.05) u_{sh, 93J}^3 \sqrt{n_{93J}}}\right) \nonumber \\
&&\times E_{PeV} \ t_{days}^{1.17} \ .
\end{eqnarray}
 Here we have assumed that the CR distribution scales as $p^{-4}$ over
 more than 6 orders of magnitude producing
 $\phi=\ln(p_{max}/p_{inj})=15$ and that $\xi_{CR}=0.05$ namely that
 5\% of the fluid kinetic energy is imparted into energetic
 particles. Here also, $p_{max}$ ($p_{inj}$) is the maximum (the
 injected) particle momentum. The growth timescale has to be shorter
 than the advection timescale towards the shock $\tau_{adv} =
 \kappa/V_{sh}^2$. The advection timescale is calculated for a
 coefficient $\kappa = \eta \kappa_{B} > \kappa_B$ taken in the
 background magnetic field (Eq \ref{Eq:Beq}) and $\kappa_B=c
 r_L/3$. We find:
\begin{equation}
\tau_{adv} = [0.24~\rm{day}] \times \eta \  E_{PeV} \times t_{day}^{1.17} \ . 
\end{equation}
The condition $\tau_{NR-st} < \tau_{adv}$ is necessary for the
instability to develop and for CR being confined around the shock, but
it is not sufficient as the magnetic fluctuations produced by the
instability uncovered by \citet{Bell04} produces small scale
perturbations. The wave number corresponding to the maximum growth
rate is \citep{Amato09} $k_{Gmax} r_{Lmax} \simeq 4 \times 10^6$ for
the conditions prevailing in SN 1993J. The acceleration and
confinement of energetic particles up to a few PeV requires magnetic
fluctuations to be generated at resonant scales $k r_L \simeq 1$, as
the NR growth rate scales as $k^{1/2}$ the magnetic field at the scale
of interest grows over a timescales about 2000 times larger than the
one obtained in Eq. \ref{Eq:TNR}. It should be noted that if the
background magnetic field is purely toroidal we may expect $\eta <
1$. In that case, the instability may not have time to develop. There
are no measurements that point towards a particular topology of the
magnetic field in the wind of different type of evolved stars
\cite{Vlemmings07}. The winds are themselves very likely
inhomogeneous, subject to turbulent motions that can introduce some
tangling effects in the background magnetic field even if on the scale
of the wind it verifies Eq \ref{Eq:Beq}. We postpone this issue for
further work and admit $\eta \ge 1$ in this work.\\

Longer wavelengths can be generated by different means. The R
instability can build up over the NR one as demonstrated in
\cite{Pelletier+06} with a ratio of the two magnetic energies reaching
$\sqrt{\xi_{CR} c/V_{sh}} \sim 0.95$ in our case. The growth timescale
for the R instability is however longer than the rate given by
Eq. \ref{Eq:TNR}. We have (see \cite{Amato09}) $\tau_{R-st} \simeq
\sqrt{\pi \sigma/8}/r_{Lmax}$ with $\sigma\sim 3 \times 10^{16}
\rm{cm^2/s^2}$ in the conditions that prevail for SN 1993J. This leads
to a growth rate $\tau_{R-st} \simeq 16 ~ \tau_{NR-st}$ at 1
PeV. Other long wavelength instabilities have to be considered. The NR
instability can be driven to non-linear stages and produce large wave
numbers. A typical timescale of non-linear saturation of the magnetic
field is about $5 \times \tau_{NR,st}$ \cite{Bell13}. Recently
\cite{Bykov+11} proposed a ponderomotive instability that builds up on
the magnetic fluctuations by the NR streaming instability. We can
evaluate the growth time-scale of such long-wavelength modes:
 \begin{eqnarray}
\tau_{LW} &=& [0.29~\rm{day}] \times \sqrt{\left({\phi/15 \over \xi_{CR}/0.05}\right)} \nonumber \\ 
&& \times {1 \over \sqrt{u_{sh, 93J}^3 A_{10}}} \times E_{PeV} t_{days} \ .
\end{eqnarray}
The parameter $A= B_{NR}^2/B_0^2 > 1$ is the level of magnetic energy
produced by the small scale instability with respect to the background
CSM and is normalized to 10 (which is rather underestimated in our
case). This timescale is sufficiently short to produce long
wavelengths on timescales shorter than $\tau_{adv}$ for $k r_{L,max}
\simeq 1$.

\subsubsection{Turbulently driven instabilities}
The stellar wind of RGB star is subject to strong fluid instabilities
and is likely turbulent \citep{Dwarkadas07}. Turbulent density and
magnetic fluctuations can have a strong impact over the SN shock and
lead to magnetic field amplification \citep{Beresnyak+09,
  Giacalone07}. Note finally that in the latter work the presence of
CR is not even necessary to produce the magnetic field
amplification. In that case the magnetic field growth time is
controlled by the coherence length of the turbulent spectrum L and by
the fluid velocity $u_{sh}$. In the conditions of SN1993J, $L/u_{sh}
\sim [0.3~\rm{year}] \ L_{0.01 pc}/u_{sh,93J}$ hence the magnetic
field has to grow over a fraction of $10^{-2}$ of this timescale.

\subsection{Maximum cosmic ray energies}
\label{S:Max}
At very early timescales after the outburst, the likely maximum energy
limitation is provided by the SNR age. The maximum energy is hence
fixed balancing the age with the acceleration time $\tau_{acc} = g(r)
\kappa_u/u_{sh}^2$, with $g(r)$ can be expressed with the shock
compression ratio as $g(r)=3r/(r-1) \times (\kappa_d/\kappa_u r
+1)$. The ratio of the down- to upstream diffusion coefficient depends
on the magnetic field orientation with respect to shock (1 in a
parallel shock, $1/r$ in a perpendicular shock). As discussed above we
admit $\kappa_d=\kappa_u/\sqrt{11}$ in our case. This corresponds to a
completely tangled magnetic field whose tangential component is
compressed by a factor 4 (see \cite{Marcowith10}).  We obtain a
maximum energy expressed in PeV units:
\begin{equation}
E_{max,age,PeV} \simeq  {12.3 \over \eta g(r)} \times (1-t_{day}^{-0.17}) \ .
\end{equation}
But rapidly (see \S \ref{Su:MFA}) the streaming instability amplifies
the magnetic field and non-linear process produces a magnetic field at
saturation. The typical saturation value expected from the NR
instability (see \cite{Bell04}) is:
\begin{equation}
\label{Eq:Bsat}
B_{sat}= [16~\rm{Gauss}] \times \sqrt{\xi_{CR}/0.05 \over \phi/15} \times t_{days}^{-1} \ .
\end{equation}
This value, as already remarked by T09, is in agreement (within a
factor of 2) with the magnetic field derived in the upstream medium
from Eq.\ref{Eq:MFo} using a compression ratio $r=4$. If only the NR
instability is at work building the magnetic field hence the maximum
particle energy is fixed by a condition over the CR areal charge
\cite{Bell13} that produces $\int \tau_{NR-st}^{-1} dt = 6.8$ (a little bit
larger than what Bell et al. have considered). The latter value corresponds 
to the amplification of the equipartition magnetic field to the value deduced 
from radio observations. In that case:
\begin{equation}
E_{max,NR,PeV} \sim 1 \times t_{day}^{-0.17} \ .
\end{equation}
But if long wavelengths fluctuations are produced by the mean of the
ponderomotive instability the maximum energy is fixed by geometrical
losses. The maximum energies are obtained with a diffusion coefficient
expressed in the amplified field and compared to $\eta_{esc} R_{sh}
u_{sh}$. In order to derive a time dependence of the maximum energy
the time dependence of the amplified field has to be specified. The
non-linear regime of the ponderomotive instability has not yet been
fully explored (see however some attempts in \cite{Rogachevski12}). In
consequence, we rely on the estimate given in Eq. \ref{Eq:Bsat} to fix
such a dependence. This gives:
\begin{equation}
E_{max,LW,PeV} \sim 55 \left({\eta_{esc} \over 0.1}\right) \times t_{day}^{-0.34} \ .
\end{equation}
Apart from the question of time dependence of the saturated magnetic
field, this value is optimistic also since the magnetic field
experienced by the highest energies is likely lower than
$B_{sat}$ \cite{Zira08}. The maximum cosmic ray (proton) energy is at any given time
the minimum of the above limits. In all cases, PeV energies at least
can be reached at early timescales after the outburst. These energies
can be probed using a gamma-ray telescope sensitive above a few
hundred TeV.\\ 

In our model exposed hereafter the maximum electron energy has also
been calculated and is found to be fixed by the radiative losses (here
synchrotron) losses.

\subsection{Cosmic-ray spectral evolution}
\label{S:CRs}
We basically follow the prescription of T09 regarding the time evolution of
the CR energy content (see their Eq. 50). The proton particle spectrum is
assumed to follow a power-law with a spectral index $s = 2$ and an exponential 
cutoff at $E_{max}(t)$,  the maximum particle energy as discussed
in \S \ref{S:Max}.
In the case of secondary electrons and positrons produced in the proton-proton 
interactions, we solved a one-zone energy equation to calculate the time 
evolution of their energy distribution $N(E,t)$:
\begin{equation}
\partial_t N(E,t) + \partial_E (\dot{L}(E) N(E,t)) = Q(E,t)
\end{equation}
The term $\dot{L}$ includes the radiative (synchrotron) losses affecting the 
secondary electrons and positrons in the the post-shock region.

\section{Multi-wavelength and multi-messenger spectra}
\label{S:Mul}

\subsection{Gamma-Ray production}
\label{S:GRa}
In this section we derive the multi-wavelength time dependent spectra
to be expected in the context of MFA driven by the instabilities
discussed in section \S \ref{Su:MFA}. First, we restricted our
analysis solely to the proton-proton interactions. Inverse Compton or
bremsstrahlung radiation have been found to be negligible in the
GeV-TeV gamma-ray domain explored here. Especially the Inverse Compton
process is highly disfavored due to the strong magnetic field at the 
forward shock. Gamma-rays once produced can be absorbed by different 
soft photon fields to produce electron-positron pairs. The main photon source 
is the SN photosphere which has been well described in the case of SN 1993J 
by \cite{Lewis94}.

\subsubsection{Gamma-gamma absorption}
The gamma-gamma absorption process is complicated by the fact that the
soft photon distribution produced by supernova photosphere looks more
and more anisotropic by the gamma-rays generated at the forward shock
as time goes on. We have performed a full calculation of the
gamma-gamma opacity $\tau_{\gamma\gamma}$ including the geometrical
effects due to the anisotropic interaction (Renaud et al. 2014 in
prep). The final gamma-ray flux is hence the unabsorbed flux $F_{\nu,
  un}$ times an attenuation factor
$\exp(-\tau_{\gamma\gamma}(E_{\gamma}))$. The gamma-gamma absorption
effect is strong just after the outburst as the interaction in nearly
isotropic and matches the derivation proposed in T09, but thereafter
drops as the ratio of the forward shock radius to the photosphere
radius reaches $\sim 3$, which happens after $\sim 5$ days. The
anisotropic absorption then boosts the gamma-ray signal one may expect
with respect to the fully isotropic case considered by T09.

\subsubsection{Cherenkov Telescope Array and H.E.S.S. detectability}
\begin{figure}
\label{F:Gam}
\includegraphics[width=0.45\textwidth]{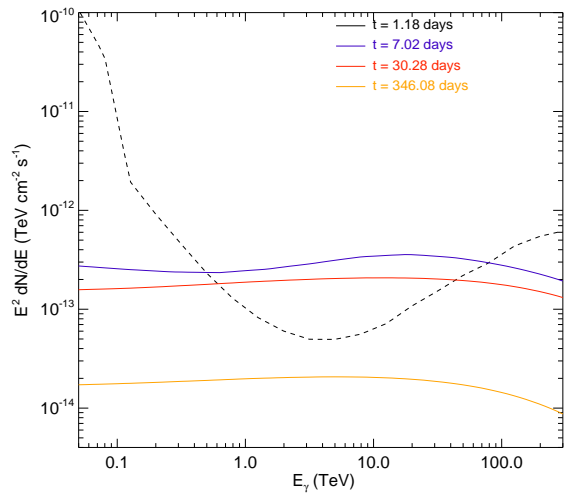}
\caption{Time dependent spectra in the CTA energy domain at four
  different times after the outburst. CTA sensitivity in 50h is shown
  with dotted lines. At t=1.18 days, the source is optically thick to
  gamma rays, which leads to a very low flux not seen in the Figure.}
\end{figure}

The time dependent gamma-ray spectra in the very-high energy (100 GeV $<$ E
$<$ 100 TeV) gamma-ray domain is displayed in figure \ref{F:Gam} (notice that the figures will be in colors only in the online version, 
whereas they will be b/w in print.)
We find that a source like SN 1993J would be easily (resp. marginally) 
detected by CTA (resp. H.E.S.S.), in particular above 1 TeV, in 20 h of 
observing time. We predict that in this type of object the best
time window to detect a gamma-ray signal is between a week and a month after 
the outburst. Prior to a week the source is optically thick to gamma-rays, 
but the gamma-gamma opacity decreases rapidly due to anisotropic effects. 
After a month the gamma-ray signal becomes too faint to be detected as the 
forward shock moves into a less dense wind medium.

\subsection{Radiation by secondaries}
\label{S:MWs}
\begin{figure}
\label{F:Sec}
\includegraphics[width=0.45\textwidth]{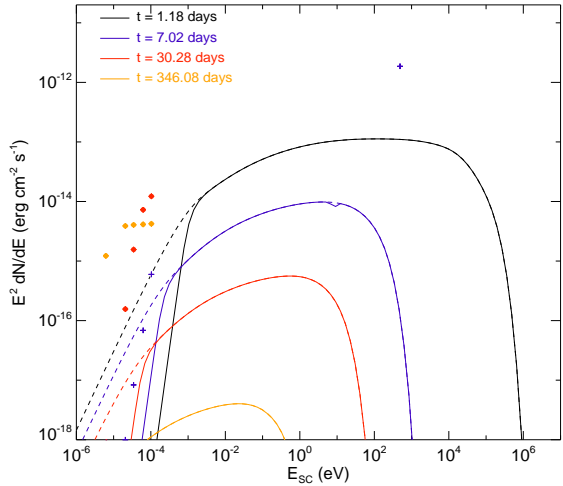}
\caption{Time dependent spectra from electron-positron secondaries at
  fourth different times after the outburst. Radio data are from
  \cite{Weiler07}. X-ray data are from \cite{Zimmermann94} and have
  been taken at about 7 days after the outburst. The dotted lines represent
  the un-absorbed synchrotron spectra while the continuous lines
  include synchrotron self-absorption (see T09).}
\end{figure}

The emission produced by the secondary particles issued from the pion
production process is displayed in Figure \ref{F:Sec}. Note that the
radiation of the electron-positron pairs produced in the
electromagnetic cascade triggered by the gamma-rays is not included in
the calculations, but this can only modify the secondary flux within
the first days after the outburst when the source is optically thick
to gamma-rays.
We find that the flux produced by the secondaries is below the level
of the observed radio and X-ray data, consistent with our previous
expectation.
\subsection{Neutrino signal}
\label{S:NEu}
\begin{figure}
\label{F:Neu}
\includegraphics[width=0.45\textwidth]{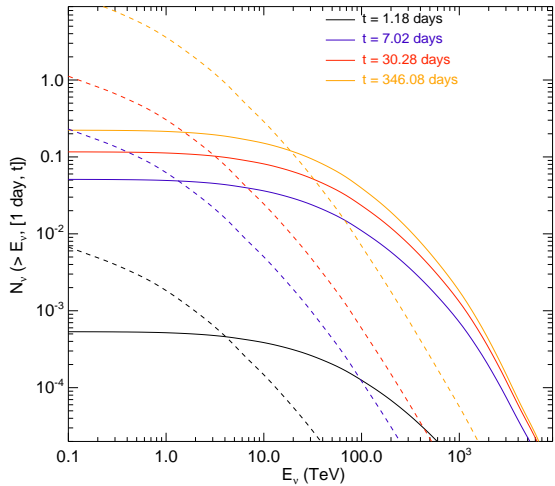}
\caption{Time dependent neutrinos flux above an energy E expected by
  an instrument equivalent to KM3NeT (continuous lines) at fourth
  different times after the outburst. In dotted lines are displayed
  the atmospheric neutrino backgrounds.}
\end{figure}

Neutrinos are also by-products of pion production. The expected flux
of neutrinos to be detected by an instrument equivalent to KM3NeT is
displayed in Figure \ref{F:Neu}.
We find that at best about 0.1 neutrinos could be expected from a SN 1993J 
type event above 10 TeV by summing the spectra between 1 and 30.28 days
after the outburst whereas for the same integration time 0.03
background neutrinos would ve expected. Hence, in order to obtain at least 
one neutrino form such a source it must be at a distance of $\sim$ 1 Mpc or 
the gamma-ray signal has to be ten times stronger.

\section{Discussion}
\label{S:Dis}
Type IIb SNe like SN 1993J are rare events. Only about 5-6\% of the
local core-collapse SNe have been classified in that category
\cite{Smartt09}. One of the most important parameter controlling the
early gamma-ray emission from SNe is the ratio of the mass loss rate
and the wind velocity $\dot{M}/v_{w}$. This parameter fixes the CSM
medium density and the different CR driven instability growth
rates. The other important parameters are: the shock velocity, the
local ionization degree of the CSM matter (see \cite{Dwarkadas14}),
the background stellar wind magnetic field and the SN peak
luminosity. The shock velocity controls the growth rate of the
instabilities and the acceleration timescale. The ionization degree is
important for the particle acceleration efficiency and may also
produces element dependent CR spectra if the ionization is incomplete
for heavier elements. The background magnetic field controls partly
the local magnetization and the shock obliquity. The SN luminosity
controls the gamma-gamma absorption process. The effects of each these
parameters deserves further investigation and will be explicitly
treated in a future work.\\

Taking the combined effects of the above parameters into account we
can examine the possibility for observations of other SN types. In
terms of high ambient density, it appears that Type IIn SNe would be
one of the most promising targets for gamma-ray telescopes. Although
high density implies lower velocity of the shock wave, the dependence
of the velocity on the density is not as large. In the Chevalier
self-similar solution, the radius of the shock wave depends on the
ambient density (or the parameter $\dot{M}/v_{w}$) only as R $\sim
({\dot{M}/v_{w}})^{-1/(n-s)}$, where $n$ is the power-law of the ejecta
density profile and $s$ is that of the circumstellar medium. For
$n=10$ and $s=2$ we get R $\sim ({\dot{M}/v_{w}})^{-1/8}$. Thus the
higher density does not lead to significantly reduced
velocities. Unfortunately these sources are even less frequent than
IIb SNe. SN IIP form the most frequent type of SN (more than 50\% of
the core-collapse SNe). These arise from RSG stars, and
should theoretically have wind mass-loss rates ranging from $10^{-7}$
to $10^{-4}$ solar masses per year. However, observationally we find
that most IIPs appear to arise from lower mass stars \cite{Smartt09}
and are less luminous in X-rays \cite{Dwarkadas14}, suggesting on
average a lower mass loss rate to wind velocity ratio. A case by case
study is mandatory in these cases before any detailed predictions can
be made. Finally, the equally rare Ib/Ic SNe are potentially
interesting as they harbor the fastest shock waves, but they are
usually associated with a Wolf-Rayet phase in the later phases of the
massive star lifetime, that have a reasonably large mass-loss rate,
but a wind velocity two orders of magnitude greater than RSGs, and
therefore a correspondingly lower wind density, which tends to delay
the peak of gamma-ray emission. Again a detail investigation of this
class is warranted.

\section{Conclusion}
\label{S:Con}
The highest energies in the CR spectrum require special environments
to be produced. In agreement with the confinement criterium multi-PeV
energies are to be either produced in large accelerators like massive
star cluster or superbubbles \cite{Bykov01, Parizot04, Ferrand10} or
in sites with high magnetic fields. The latter are found in young SNRs 
or in radio SNe (see T09 and \cite{Bell13}). In this work we argue that 
radio SNe could produce multi-PeV particles through the combination of 
favorable effects: fast shocks with velocities about 10\% of the speed 
of light, high density CSM as produced by the wind of RSG stars and low 
wind magnetizations. A high ionization degree of the CSM would also ease
the particle acceleration process. Under the condition that the
background magnetic field has a turbulent component that prevents it
from being purely toroidal, we have shown that the different
instabilities driven by the acceleration of energetic particles could
grow over intra-day timescales. If there is still some discussion
about the exact way the magnetic fields can saturate and about the
properties of the turbulence in the configuration of such fast shocks
one can argue that protons with energies 1-10 possibly up to 50 PeV
can be accelerated within a few weeks after the SNe outburst. This
model is also consistent with the magnetic field strengths deduced
from the radio observations of the most luminous objects like SN 1993J
(T09). Considering the fiducial case of SN 1993J, we have shown using
a simple acceleration one-zone model that gamma-rays produced by pion
decay could be easily detectable by CTA once the source become optically 
thin to pair production, i.e. about 7 days after the outburst. The 
predicted high-energy neutrino flux above 10 TeV from a SN 1993J type event
is only about 0.1 neutrino for an instrument like Km3NeT. We have checked 
that the radio and X-ray fluxes from the secondaries are consistent with the 
data. Our scenario suggests the possibility to produce clear predictions and 
hence to be testable by the forthcoming high-energy instruments. We have also 
discussed other types of SNe, and identified IIn SNe as possible strong 
gamma-ray emitters. But a full investigation of this possibility deserves 
further work. This work investigates a small parameter space of the SN forward 
shock properties and hence cannot address the important question of the
formation of the CR spectrum (see the discussion in \cite{Ptuskin10}). This 
aspect will also be addressed in a forthcoming work.







\end{document}